\documentclass[a4paper]{amsart}
\usepackage{amssymb}

\newtheorem{thm}{THEOREM}[section]

\theoremstyle{remark}
\newtheorem{rem}[thm]{REMARK}

\numberwithin{equation}{section}

\newcommand{\infspec}{{\rm inf\ spec\ }}
\newcommand{\R}{{\mathbb R}}

\newcommand{\C}{{\mathbb C}}
\newcommand{\Ow}{{\mathcal O}}
\newcommand{\ora}{{|0\rangle}}

\newcommand{\Da}{{D}^\ast}
\newcommand{\mA}{{\mathcal A}}
\newcommand{\err}{{\mathcal Err}}
\newcommand{\Ea}{{E}^\ast}
\newcommand{\F}{{\mathcal F}}
\newcommand{\Em}{{\mathcal E}}
\newcommand{\Ll}{{\mathcal L}}

\newcommand{\Hh}{{\mathcal H}}
\newcommand{\eps}{\varepsilon}

\newcommand{\aan}{a_{\lambda}}
\newcommand{\ac}{a^{\ast}_{\lambda}}

\newcommand{\ean}{\varepsilon_{\lambda}}
\newcommand{\half}{\mbox{$\frac{1}{2}$}}

\newcommand{\al}{{\alpha}}
\newcommand{\pa}{{\parallel}}

\newcommand{\so}{{\Sigma_\alpha}}
\newcommand{\Eal}{{\mathbf{E}_\alpha}}
\newcommand{\vs}{\vec \sigma}

\newcommand{\as}{\sqrt{\alpha}}

\newcommand{\pb}{{\bar \phi}}
\newcommand{\la}{\Lambda}

\newcommand{\ua}{\uparrow}
\newcommand{\hpt}{{\widetilde{\psi_n}}}
\newcommand{\hp}{ \overline{\psi_n}}
\newcommand{\pe}{\psi_{n+1}}
\newcommand{\hps}{ \psi_{n+2}}

\newcommand{\bh}{{\bf H_\al}}
\newcommand{\be}{\beta}

\begin{document}

\title[Increase of the binding energy of an electron]{Increase of the binding energy of an electron by coupling to a photon field}
\author{ Christian Hainzl}
\address{Mathematisches Institut, LMU M\"unchen,
Theresienstrasse 39, 80333 M\"unchen, Germany}
\email{hainzl@mathematik.uni-muenchen.de}

\thanks{The author has been supported by a Marie Curie Fellowship
of the European Community programme \lq\lq Improving Human Research Potential and the
Socio-economic Knowledge Base\rq\rq\ under contract number HPMFCT-2000-00660.}

\date{April 28, 2002}

\begin{abstract}
We look at an electron in the field of an arbitrary external potential
$V$, such that the Schr\"odinger operator $p^2 + V$ has at least
one eigenvalue, and show that by coupling to a quantized radiation
field the binding energy increases, at least for small enough
values of the coupling constant $\alpha$. Moreover, we provide 
concrete numbers for $\al$, the ultraviolet cut-off $\la$, and the radiative correction for which our procedure works.
\end{abstract}

\maketitle

\section{INTRODUCTION}

Quantum Electrodynamics (QED) has proved one of the most successful theories in physics.
Nevertheless the treatment by means of perturbation theory and the renormalization procedure
has caused some uneasiness among mathematicians.

A couple of years ago Bach, Fr\"ohlich, and Sigal initiated a rigorous non-perturba\-tive study
of non-relativistic QED. Among other results they proved in \cite{BFS} the existence of a ground state
for small values of the coupling parameter $\alpha$. At least for one particle
Griesemer, Lieb, and Loss \cite{GLL}, which we refer to for an extensive list of references, 
succeeded in removing these restrictions, whereas for more particles the analogous result 
is shown under very reasonable conditions.

Based on this result in \cite{GLL},
the author, V. Vougalter, and S. Vugalter recently proved in \cite{HVV}  
that coupling a charged boson to a radiation field {\it enhances binding}.
Namely, for some short-ranged potential $V$ binding can occur although
the operator $p^2 + V$ does not have a negative eigenstate. Intuitively this phenomenon 
does not surprise, because the \lq\lq photon cloud\rq\rq \ surrounding the particle 
should increase its effective mass which lowers the ground state energy. 
The result in \cite{HVV} is valid in the case the parameter $\al$, which measures the coupling to 
the field,
is small enough. This restriction stems from the fact that we are only able to control the
self-energy of a particle, $\so$, which is a complicated function on $\al$ and the ultraviolet cut-off $\la$
(cf. \cite{LL}), for small values of $\al$.

In the present paper we study the ground state energy of one electron
moving in an arbitrary external potential. The only requirements needed are the self-adjointness of $p^2 + V$ 
and the existence of a ground state. We couple this operator to a 
quantized radiation field through
the Pauli-Fierz Hamiltonian, $H_\al=T+V$, with $T = [\vs \cdot (p+ \as A)]^2 + H_f$.
Our main theorem  states that the {\it binding energy},
\begin{equation}\label{ben}
\Delta E= \infspec T -\infspec H_\al,
\end{equation}
increases as soon as the radiation ($\al >0$) is turned on.
Our result again holds for sufficiently small values of $\al$.

Although  not necessary for our main result concerning the binding energy defined by (\ref{ben}),
we assume the external potential $V$ to satisfy analogous conditions as in \cite{GLL}, which are physically very reasonable, in order 
to guarantee 
that the $\infspec H_\al$ is a ground state eigenvalue. 

Our considerations concerning  the self-energy of an electron are based on \cite{H1}. Nevertheless, 
referring to the calculations in \cite[Theorem 1]{H1},
we generalize
the notation and improve the error estimates in $\al$.

In Section \ref{shift} we evaluate the radiative correction in the case of a nucleus  
and we provide concrete numbers for $\al $ and $\la$ 
for which enhanced binding can be guaranteed.  Furthermore, we comment on non-perturbative mass
renormalization.

\section{DEFINITIONS AND MAIN RESULTS}

The self-energy of an electron is described by the Pauli-Fierz
operator
\begin{equation}\label{rpf}
T = (p + \sqrt{\al}A(x))^2 + \as\vs\cdot B(x) + H_f.
\end{equation}

The unit of energy is $mc^2/2$, where $m$ is the electron mass,
the unit length is $l_c = 2 \hbar /mc$, twice the Compton
wavelength of the particle, and $\al= e^2/\hbar c$ is the {\it
fine structure constant}, where $e$ is the charge of the electron.
The physical value of $\al$ is $\sim 1/137$. In the present paper
$\al$ plays the role of a small, dimensionless number, which measures 
the strength of the coupling. Our results
hold for {\it sufficiently small values } of $\al$. $\vs$ is the
vector of Pauli matrices $(\sigma_1,\sigma_2,\sigma_3)$. Recall,
the $\sigma_i$ are hermitian $2\times 2$  matrices and fulfill the
anti-commutation relations $\sigma_i\sigma_j + \sigma_j\sigma_i
=2I_{\C^2} \delta_{i,j}$. The operator $p= -i\nabla$, while $A$ is
the magnetic vector potential. The magnetic field is $B = {\rm
curl} \ A$. The unit of $A(x)^2$ is $mc^2/2l_c$. We fix the gauge
${\rm div} A =0$. The units are chosen as in \cite{GLL}.

The underlying Hilbert space is
\begin{equation*}
\Hh = \Ll^{2}({\mathbb {R}}^{3};\C^2)\otimes \F
\end{equation*}
where $\F$ is the Fock space for the photon field.

The vector potential is
\begin{equation*}
A(x) = \sum_{\lambda = 1,2} \int_{\R^3}
\frac{\chi(|k|)}{|k|^{1/2}} [\ean(k) \aan(k) e^{ikx} + \ean
(k)\ac(k) e^{-ikx}] dk = D(x) + D^\ast (x),
\end{equation*}
and  the corresponding magnetic field reads
\begin{eqnarray}\nonumber
B(x) &=& \sum_{\lambda = 1,2} \int_{\R^3}
\frac{\chi(|k|)}{|k|^{1/2}} [(k\times i\ean(k)) \aan(k) e^{ikx} +
\overline{(k \times i\ean(k))} \ac(k) e^{-ikx}] dk \\ \nonumber &=& E(x) + E^\ast
(x),
\end{eqnarray}
where the operators $\aan$ and  $\ac$ satisfy the usual
commutation relations
\begin{equation*}
[a_\nu (k),\ac(q)] = \delta(k-q)\delta_{\lambda,\nu}, \quad
[\aan(k), a_\nu(q)] = 0, \ \ {\rm etc}.
\end{equation*}
The vectors $\ean(k) \in \R^3$ are any two possible orthonormal
polarization vectors perpendicular to $k$.

The function $\chi(|k|)$ describes the ultraviolet cutoff on the
wavenumbers $k$. We choose  $\chi$  to be the Heaviside
function $\Theta(\Lambda -|k|)$. Throughout the paper we assume
$\la$ to be an arbitrary but fixed positive number. The {\it physical}
value of $\la $ is expected to be less or equal $ 2$ which corresponds to
the energy $mc^2$. This energy is a natural upper bound for the
cut-off parameter since it is the maximum value that guarantees
that no pair-production takes place.

The photon field energy $H_f$ is given by
\begin{equation*}
H_f = \sum_{\lambda= 1,2} \int_{\R^3} |k| \ac (k) \aan (k) dk.
\end{equation*}

Our first theorem concerns the behavior of the self-energy of 
an electron
\begin{equation}
\so = \infspec T,
\end{equation}
for fixed $\la$ and small $\al$. Although it is mainly based on \cite[Theorem 1]{H1}, we give 
a slightly different proof including an improvement of the error estimate.

Analogously to \cite[Theorem 1]{H1} we first define the critical cut-off 
parameter
\begin{equation}
\bar \la = \mbox{Max}\{\la | (f,E \mA^{-1} \Ea f) \leq \pa f\pa^2 \langle 0| E \mA^{-1} \Ea | 0\rangle \, \forall f \in \Ll^2(\R^3)\},
\end{equation}
where $\ora $ is the vacuum in the Fock space $\F$.
\begin{thm}\label{t1}
Let $\la \leq \bar \la$. Then 
\begin{equation}\label{tq1}
|\so - 8\pi \al [\la - \ln(1+\la)]| \leq C \al^2,
\end{equation}
for some constant $C > 0$ depending on $\la$.

For $\la > \bar \la$ (\ref{tq1}) holds with an error $\Ow(\al^{5/4})$.
\end{thm}
\begin{rem}
Fortunately, $\bar \la$ can be evaluated explicitly and is about $\bar \la \sim 12.6$.
\end{rem}
We consider an electron in an arbitrary external potential,
described by a real valued function  $V$, independent of $\al$, which satisfies  (1) that
 the negative part of $V$, $V_-$,  is dominated by the kinetic energy  $p^2 =
-\Delta$,
\begin{equation}
V_- \leq -\eps \Delta + C_\eps,
\end{equation}
for any $\eps >0$ and for some positive constant $C_\eps$,
such that the self-adjointness and boundedness from below of $p^2 + V$ is guaranteed.
Additionally, since we want the essential spectrum to start at $0$, we require (2) that 
$V_-$ tends to $0$ at infinity.
Furthermore, we assume (3) that
the operator $p^2 +V$ has at least
one negative energy bound state.

The corresponding operator with a quantized radiation field reads
\begin{equation}
{\bf H_\al}= T + V.
\end{equation}
Recently, Griesemer, Lieb, and Loss \cite{GLL} proved for all
values of $\al, \la$ that $\bh$ has an eigenstate, as well as that
the binding energy $ \so - \Eal$, with
\begin{equation}
\Eal = \infspec \bh,
\end{equation}
cannot decrease, i.e.
\begin{equation}\label{bc}
\Eal - \so\leq -e_0,
\end{equation}
where $-e_0 = \infspec [p^2 + V]$. As the main result we show that
at least for small values of the coupling constant $\al$ the
binding energy increases, namely (\ref{bc}) holds with strict
inequality.

\begin{thm}\label{t2}
Let $\la \leq \bar \la$. Let $V$ satisfy the conditions (1)-(3), and 
$\phi$ be the ground state of $p^2 + V$ with corresponding energy $-e_0$. Then
\begin{equation}\label{es}
\Eal - \so \leq -e_0 - \al \Em(V,\la) + \Ow(\al^2)
\end{equation}
for some positive number  $\Em(V,\la)$ depending on $ V$ and the cut-off $\la$.
Moreover,  there exists a number $\rho > 0$, such that for all $\al
\in (0,\rho]$
\begin{equation}
\Eal - \so < -e_0.
\end{equation}
For $\la > \bar \la$ (\ref{es}) holds with an error $\Ow ( \al^{5/4})$.
\end{thm}
\begin{rem}
The equation (\ref{es}) holds for all real $V$, such that  $p^2 + V$ is self-adjoint and has  a ground state.
But only for $V$ satisfying $(1) -(3)$ we know that $H_\al$ has a ground state.

We actually check  that  $\Em(V,\la)$ is bounded from above by
\begin{equation}
\Em(V,\la) \leq \pa p \phi\pa^2 \frac{32\pi}3 \ln[1+\la].
\end{equation}
\end{rem}
Moreover, we strongly conjecture that apart from errors of order $o(\al)$ Equation (\ref{es}) holds with
equality.

\section{PROOF OF THEOREM \ref{t1}}

\subsection{Upper bound for $\so$}

We define the  sequence of trial wave functions
\begin{equation}\label{tf}
\Psi_n = \{f_n \cdot \ua, -\as  [p_{x-\xi}^2 + H_f]^{-1}(\vs \ua)
\cdot\Ea f_n , 0,0,\ldots \},
\end{equation}
with $f_n \in \Ll^2(\R^3)$, $\pa f_n\pa=1$ and $\pa p f_n \pa \to
0$ for $n\to \infty$. Notice,
\begin{equation}
\Ea  f_n(x,x-\xi) \equiv f_n(x)
\otimes [\Ea (x)|0\rangle] (x-\xi)
\end{equation}
is a function depending on
$(x,x-\xi)$. 
The function $[\Ea(x)\ora] (x-\xi)$ is intensively described in \cite[Section 2]{H1}.
Recall, $\ua $ denotes the spin up vector, $x$ the position of the electron  and $\xi$ 
is the photon variable. For convenience we introduce 
the relative 
variable  $\eta = x
-\xi$, and  we keep in mind
\begin{equation}
\left([p^2 + H_f]\Ea f_n\right)(x,\eta) = ([p_x + p_\eta]^2 + |p_\eta|)\Ea f_n(x,\eta).
\end{equation}

For our calculations we just need to know that the Fourier transform of $\Ea \ora$ is
\begin{equation}
\F[\Ea\ora (\eta)](k) \equiv  H(k) = \sum_{\lambda=1,2} H_\lambda (k)= i\sum_{\lambda=1,2}  (k\times
\eps_\lambda(k))\frac {\chi(|k|)}{|k|^{1/2}}
\end{equation}
(cf. \cite{H1}). Denote the 1-photon part of $\Psi_n$ as $\psi_n^1$. Notice,
\begin{equation}
(\psi_n^1, p^2\psi_n^1) = ( \psi_n^1, [p_x + p_\eta]^2 \psi_n^1) =
(\psi_n^1, p_\eta^2 \psi_n^1) + \mbox{ Error}(\psi_n^1),
\end{equation}
where, due to our assumption that $\pa p f_n\pa$ tends  to $0$, ${\rm Error}(\psi_n^1) \to 0$ as $n \to \infty$.

Furthermore, we compute
\begin{eqnarray}\nonumber
&& (\psi_n^1, [p_\eta^2 + H_f] \psi_n^1) + 2\as \Re \left( (\vs
\ua)\cdot \Ea f_n,\psi_n^1\right) = \\ \nonumber && -\al
\left( (\vs \ua) \Ea f_n, [p_\eta^2 + H_f]^{-1} (\vs \ua) \cdot \Ea
f_n\right) = -\al \pa f_n\pa^2 \langle 0|E [p^2 + H_f]^{-1}\Ea\ora.
\end{eqnarray}
Evaluations analogously to  \cite[Section 3.1]{H1}  show
\begin{eqnarray}\nonumber
\langle 0|E [p^2 + H_f]^{-1}\Ea\ora &=&\sum_{\lambda =1,2} \int \frac{| H_\lambda(k)|^2}{|k|^2
+ |k|} dk \\ &= & 2\int \frac{|k|\chi(|k|)}{|k|^2
+ |k|} dk = 4\pi[\la^2 - 2[\la - \ln(1 + \la)]].
\end{eqnarray}
Therefore, we arrive at
\begin{eqnarray}\nonumber
\so & \leq& \lim_{n\to \infty} (\Psi_n,T\Psi_n)/(\Psi_n,\Psi_n)  = 4\pi \al \la^2 - \al
\langle 0|E [p^2 + H_f]^{-1}\Ea\ora + \Ow(\al^2)\\ \nonumber
&\leq& 8\pi \al [\la - \ln(1+\la)] + \Ow(\al^2),
\end{eqnarray}
where
\begin{equation}
\Ow(\al^2) \leq \al^2 \langle 0|E [p^2 + H_f]^{-1}\Ea\ora \pa [p^2 +
H_f]^{-1} \Ea \ora\pa^2.
\end{equation}

\subsection{Lower bound on $\so$.}

We take some $\Psi \in \Hh$
close to $\so$, i.e. $(\Psi,T\Psi) \leq 8\pi\al \Lambda  + O(\al^2)$,
which is obviously possible and $\pa \Psi\pa =1$. Since $T \geq H_f$
we immediately get for those $\Psi$'s
\begin{equation}\label{aprih}
(\Psi, H_f\Psi) \leq 8\pi\al \Lambda  .
\end{equation}
Therefore, by means of Schwarz' inequality we obtain,
as in \cite{H1},
\begin{eqnarray}\label{39}
2\as(\Psi,pA\Psi) &\leq& a\pa p\Psi\pa^2 + 4a^{-1}\al \pa D\Psi\pa^2\\ \label{310}
\as(\Psi,\vs  B\Psi) &\leq& c \la^3 \al \pa \Psi\pa^2 + c^{-1} (1/\la^3) \pa E \Psi\pa^2,
\end{eqnarray}
for any $a,c > 0$.
With \cite[Lemma A. 4]{GLL},
\begin{equation}\label{energy}
\Da D \leq 8\pi \la H_f, \quad \Ea E \leq \frac {8\pi}3 \la^3 H_f,
\end{equation}
we infer 
\begin{equation}\label{aprip}
\pa p\Psi\pa^2 \leq C_1\al \la^3 +C_2\al \la
\end{equation}
for some constants $C_{1,2}>0$.
Equations (\ref{aprih}) and (\ref{aprip}) will be decisive to
control our error estimates. Recall, 
\begin{equation}
A^2 = 4\pi \al \la^2 + 2 \Da D +  2 \Re DD.
\end{equation}
Thus, 
\begin{equation*}
(\Psi,T\Psi) \geq 4\pi\al \Lambda^2 \pa \Psi \pa^2 + \Em_0[\psi_0,\psi_1] + \sum_{n =0}^{\infty}
\Em[\psi_n,\psi_{n+1},\psi_{n+2}]
+2\al (\Psi,D^\ast D\Psi) ,
\end{equation*}
where
\begin{equation}
\Em_0 [\psi_0,\psi_1] = ( \psi_1, \mA \psi_1)  +
2\as{\Re} (\vs \Ea \psi_0,\psi_1)
\end{equation}
and 
\begin{eqnarray}\nonumber
&&\Em[\psi_n,\psi_{n+1},\psi_{n+2}] = (\psi_{n+2} , \mA \psi_{n+2}) +\\ &&\,\,\quad + 2\Re\left([\as \vs \Ea + 2\as \Da p]\psi_{n+1}
+ \al \Da\Da \psi_n, \psi_{n+2}\right),
\end{eqnarray}
where
\begin{equation}
\mA= p^2 + H_f.
\end{equation}
Next, we proceed analogously to \cite{H1}.
We denote a general $\Psi \in \Hh$ by
\begin{equation}
\Psi = \{\psi_0 (x), \psi_1(x,\eta_1),....,\psi_n(x,\eta_1,...,\eta_n),...\},
\end{equation}
with $x-\xi_i=\eta_i$, where we omit for simplicity the photon polarization variables $\lambda_i$.
For convenience we will work in momentum representation
\begin{equation}
\F[\psi_n(x,\eta_1,...,\eta_n)](l,k) = \psi_n(l,k),
\end{equation}
with $k = (k_1,...,k_n)$.

First of all, we follow \cite[Section 4.2]{H1}   and decompose $\psi_n$ into a part where $
\sum_{i=1}^{n} |k_i| \leq \Lambda$
and into its complement.
This can be done by an additional cut-off function
$\chi_\Lambda (k) = \Theta(\Lambda - \sum_{i=1}^{n} |k_i| ).$
Respectively, we denote
\begin{equation*}
\psi_n(l,k) \chi_\Lambda (k) = \widetilde{ \psi_n} (l,k), \quad \psi_n(l,k)
[1-\chi_\Lambda (k) ]= \hp(l,k).
\end{equation*}
Therefore, we rewrite
\begin{eqnarray}\nonumber
&&\Em[\psi_n,\psi_{n+1},\psi_{n+2}] = \Em[\hpt,\psi_{n+1}, \hps] \\ && + 2\al[ \Re (D^\ast \hp , D\hps) +
(D\hps,D\hps)].
\end{eqnarray}
Using first Schwarz' inequality and afterwards the relations  (\ref{energy})
we derive
\begin{eqnarray}\nonumber
&& 2\al[ \Re (D^\ast \hp , D\hps) + (D\hps,D\hps)] \geq -\frac{\al}{2}
(D^\ast\hp,D^\ast\hp)\geq\\ \label{rose}
&& \geq  -\frac{\al}{2} [\Lambda^2\pa \hp\pa^2 + \Lambda
(\hp,H_f \hp)]\geq - \Ow(\al  )\Lambda (\hp,H_f \hp),
\end{eqnarray}
for, from the definition of $\chi_\Lambda(k)$, 
\begin{equation*}
(\hp,H_f \hp) \equiv \sum_{i=1}^{n} \int \hp(l,k)^2 |k_i|  \geq \Lambda
\pa \hp \pa^2.
\end{equation*}
Next, we regard the term $\Em[\hpt,\psi_{n+1},\hps]$.
By Schwarz' inequality we easily compute
\begin{equation}\label{schw}
\Em[\hpt, \psi_{n+1}, \hps]\geq - \pa \as [\mA^{-1/2} \vs \Ea + \mA^{-1/2} \Da p ]\psi_{n+1}
+\al \mA^{-1/2} \Da \Da \hpt \pa^2.
\end{equation}
Similarly,
\begin{equation}\label{eo}
\Em_0[\psi_0,\psi_1] \geq -\al (\vs \Ea \psi_0, \mA^{-1} \vs \Ea \psi_0).
\end{equation}
Now we are going to evaluate the r.h.s. of (\ref{schw}).
First, we consider the diagonal terms.
The most important one is
\begin{equation}
-\al (\vs \Ea \pe, \mA^{-1} \vs \Ea \pe),
\end{equation}
which, due to the well known
anti-commutation relations
\begin{equation}
\sigma_i\sigma_j + \sigma_j\sigma_i = 2 \delta_{i,j},
\end{equation}
is equivalent to
\begin{equation}
-\al (\pe, E\mA^{-1} \Ea\pe).
\end{equation}
Following \cite[Section 3.2]{H1} this 
term can be bounded
from  (for $\la \leq \bar \la$) below by 
\begin{equation}\label{smc}
-\al \pa \pe\pa^2 \langle 0| E\mA^{-1} \Ea \ora - \Ow(\al) \la^2  (\pe, H_f \pe),
\end{equation}
which is also valid for (\ref{eo}); namely,
\begin{equation}
-\al (\vs \Ea \psi_0, \mA^{-1} \vs \Ea \psi_0) \geq -\al\pa \psi_0\pa^2 \langle 0| E\mA^{-1} \Ea \ora.
\end{equation}
Another diagonal term is 
\begin{equation}\label{diego}
-\al^2(\hpt, DD \mA^{-1} \Da\Da \hpt),
\end{equation}
which
according to  \cite[Section 4.2]{H1}  is estimated by
\begin{equation*}
(\ref{diego}) \geq - \al^2 \pa \hpt \pa^2 \langle 0| DD \mA^{-1} \Da \Da \ora
- \Ow(\al^{9/4}) \pa \pe\pa^2 + \Ow(\al)\la (\pe, H_f \pe).
\end{equation*}
To evaluate 
the third diagonal term we first introduce the notation
\begin{equation}
\F[\Da\ora](k) \equiv G(k) =\sum_{\lambda= 1,2} G_\lambda(k)= \sum_{\lambda= 1,2} \eps_\lambda(k) \frac{\chi(|k|)}{|k|^{1/2}}.
\end{equation}
Thus,
\begin{eqnarray}\label{mc}
&& -\al (p\Da \pe, \mA^{-1} p \Da \pe) = \\\nonumber&& \quad = \sum_{\lambda =1,2} -\al \left[ \int \frac{ \left[G_\lambda(p) \cdot K\right]^2 |\pe(l,k)|^2}{|K|^2 
+|p| + \sum_{i=1}^{n+1} |k_i|}dldpdk + \right. \\ \nonumber&& \quad
\left. n \int \frac{ \left[G_\lambda(p) \cdot \bar K\right] \left[G_\lambda(q) \cdot \bar K\right]\overline{\pe(l,p,\bar k)}\pe(l,q,\bar k)}{|{\bar K}|^2 +|p| + |q| 
+\sum_{i=2}^{n+1} |k_i|} dldpdqd\bar k\right],
\end{eqnarray}
with $k=(k_1,...,k_{n+1})$, $\bar k = (k_2,...,k_{n+1})$, $K=l+ p+
\sum_{i=1}^{n+1} k_i$ and $\bar K= l+ p+ q+
\sum_{i=2}^{n+1} k_i$.
The first term in the r.h.s. is simply bounded from below by $-\al 8\pi \la \pa p\pe\pa^2$.
Moreover, using 
\begin{equation}
\frac{|{\bar K}|^2}{|{\bar K}|^2  +|p| + |q| +\sum_{i=2}^{n+1} |k_i|} \leq 1,
\end{equation}
the second term is estimated by $-\al 8\pi \la^2 (\pe, H_f \pe)$.

Next we consider the off-diagonal terms.
Notice,
\begin{eqnarray}\nonumber
&&\al\Re (p\Da \pe, \mA^{-1}\vs \Ea \pe) =\\ \nonumber && \quad =\sum_{\lambda=1,2}\al\left[ \Re \int \frac{ \left[G_\lambda(p) \cdot K\right] H_\lambda(p)\cdot
 \langle \pe,\vs \pe\rangle_{\C^2}(l,k)}{|K|^2 +|p| + \sum_{i=1}^{n+1} |k_i|} dldpdk \right.\\ \nonumber
&& \quad \left.+ n \Re \int \frac{ \left[G_\lambda(p) \cdot \bar K\right] H_\lambda(q)\cdot  \langle \pe(l,p,\bar k),\vs \pe(l,q,\bar k)\rangle_{\C^2}}
{|{\bar K}|^2 +|p| + |q|+ \sum_{i=2}^{n+1} |k_i|}dldpdqd\bar k\right].
\end{eqnarray}
The first term in the r.h.s. vanishes, because the integral is purely imaginary. For the second term 
we use $\frac {|a|}{a^2 + b} \leq \half b^{-1/2}$
and bound it by 
\begin{eqnarray} \nonumber
&\leq& \al n \int \frac{|G(p)||H(q)|}{(|p| + |q|)^{1/2} |p|^{1/2} |q|^{1/2}} \sqrt{\rho_{\pe}(p) |p|}
\sqrt{\rho_{\pe}(q) |q|} dp dq\\ \nonumber
&\leq& \Ow(\al)\la^{3/2} (\pe, H_f \pe),
\end{eqnarray}
with
\begin{equation}
\rho_{\pe}(p) = \int |\pe(l,p,\bar k)|^2 d l d\bar k.
\end{equation}

Obviously, the term $2\al^{3/2} \Re(p\Da \pe, \mA^{-1} \Da\Da \hpt)$ is dominated by
\begin{equation}
\al \pa \mA^{-1/2}p\Da \psi_{n+1}\pa^2 + \al^2 (\hpt ,DD\mA^{-1} \Da\Da \hpt),
\end{equation}
where we can apply above estimates. 

Finally, it is possible to bound
$\al^{3/2}\Re (\vs \Ea \pe, \mA^{-1} \Da\Da \hpt)$ from above
by
\begin{equation*}
\Ow(\al^{3/2}) \la^{3/2}\left[(\pe, H_f \pe) + \la (\psi_n, H_f \psi_n)\right].
\end{equation*}
Collecting above estimates and summing over all $n$ we arrive at 
\begin{eqnarray}\nonumber
(\Psi, T\Psi) &\geq& 4\pi \al \la^2 \pa\Psi\pa^2 - \al\pa \Psi\pa^2 \langle 0|E A^{-1} \Ea | 0 \rangle-
\\ && - \Ow(\al)(\Psi, H_f\Psi) - \Ow(\al)\pa p\Psi\pa^2 - \Ow(\al^2) \pa \Psi \pa^2.
\end{eqnarray}
By our a priori estimates (\ref{aprih}) and (\ref{aprip}) we prove the Theorem.

The second part follows immediately from \cite[Theorem 1]{H1}.

\section{PROOF OF THEOREM \ref{t2}}

It suffices to use 
a cleverly chosen trial wave function. We furthermore
conjecture that our choice is optimal in the sense that (\ref{es}) holds as equality
apart from errors of order $\Ow(\al^{3/2})$. We assume that
$\phi(x) \in \Ll^2(\R^3)$ is the ground state of $p^2 + V$, i.e.
\begin{equation}
(p^2 + V)\phi= -e_0 \phi.
\end{equation}
Recall, 
\begin{equation*}
\Ea \phi(x, \eta) = \phi (x)\otimes [\Ea (x)|0\rangle](\eta),\quad \Da p\phi=
\sum_{i=1}^3 p_i\phi (x) \otimes [\Da_i |0\rangle](\eta)
\end{equation*}
are 1-photon functions depending on $x$ and
the relative coordinates $x-\xi = \eta$ (cf. \cite{H1}). Therefore,
in configuration space, where $\psi(x,\eta)$ denotes one of these functions, we have (cf. the previous section)
\begin{equation}
[(p^2 + H_f)\psi](x,\eta)= ([p_x + p_\eta]^2 + |p_\eta|)\psi(x,\eta).
\end{equation}
For sake of convenience we define the operator
\begin{equation}
A_V =p_x^2 + p^2_\eta +H_f +V + e_0= (p_x^2 + V
+e_0)\otimes \mathbb{I} + \mathbb{I}\otimes (p_\eta^2 + H_f),
\end{equation}
which acts on $\Ll^2\left((\R^3;\C^2)\otimes( \R^3;\C^2)\right) $ and is
obviously positive and invertible.

Now, we choose our trial wave function $\Psi \in \Hh$
\begin{equation}
\Psi = \{\pb, -2\as A_V^{-1}\Da p\pb - \as A_V^{-1} \vs \Ea \pb, 0,
0,...\},
\end{equation}
where $\pb= \phi \cdot \ua$. We
assume $\pa \phi\pa =1$ and for simplicity we denote the 1-photon
part as $\psi_1$. Notice, $\vs \Ea \pb= \phi \otimes (\vs \ua)
\cdot \Ea |0\rangle$.

First, observe that $A_V \pb = 0$ yields
\begin{equation}
A_V \vs \Ea \pb=(p_\eta^2 + H_f)  \vs \Ea \pb.
\end{equation}
Therefore, since $A_V$ and $p_\eta + H_f$ commute by definition, we infer
\begin{equation}
A_V^{-1}  \vs \Ea \pb= (p_\eta^2 + H_f)^{-1} \vs \Ea \pb,
\end{equation}
and
\begin{equation}
\left(\pb,  \vs E A_V^{-1}  \vs \Ea \pb\right) = \pa \phi\pa^2
\langle 0|E \mA^{-1}\Ea|0\rangle,
\end{equation}
with $\mA = p_\eta^2 + H_f$. Thus, we evaluate
\begin{eqnarray}\nonumber
(\Psi, \bh \Psi) &=& 4\pi \al \la^2 \pa \Psi\pa^2 - e_0 \pa
\Psi\pa^2 + (\psi_1, A_V \psi_1) + 2\as \Re \left((\vs \Ea + 2\Da
p)\pb, \psi_1\right) \\ && + 2\al(\psi_1,\Da D \psi_1) + 2
(\psi_1,p_x \cdot p_\eta \psi_1).
\end{eqnarray}
Notice, the cross term
\begin{equation}
\Re(\pb, \vs E A_V^{-1} \Da p \pb) = \Re(\pb, \vs E \mA^{-1} \Da p\pb)
\end{equation}
vanishes as in the previous section, because it is purely imaginary. Therefore, by our choice of
$\psi_1$ we get
\begin{eqnarray}\nonumber
&&(\psi_1, A_V \psi_1)+ 2\as \Re \left((\vs \Ea + 2\Da p)\pb,
\psi_1\right)= \\ && = -4\al (\Da p_x \phi, A_V^{-1}\Da p_x \phi) -
\al \pa\phi\pa^2 \langle 0| E \mA^{-1}\Ea |0\rangle,
\end{eqnarray}
which indicates that our choice of $\psi_1$ is optimal. Moreover,
we have
\begin{eqnarray} \nonumber
&&(\psi_1,p_x \cdot p_\eta \psi_1) = \al (\Da p_x \phi, A_V^{-1}
p_x\cdot p_\eta A_V^{-1} \Da p_x \phi) \\ \nonumber && - \al (\vs \Ea \pb, \mA^{-1}
p_x\cdot p_\eta \mA^{-1}\vs \Ea \pb) +2\al\Re (\Da p_x
\pb, A_V^{-1}p_x\cdot p_\eta A_V^{-1}\vs \Ea \pb).
\end{eqnarray}
The first term in the r.h.s. vanishes by integrating over the
$\eta$-variable (the best way to see it is using the
representation in momentum space). The second term vanishes when
integrating over the $x$-variable (notice, $(\phi, p_x\phi)=0$), and
the third term vanishes, because it is again purely imaginary.

Using above considerations and the fact that 
\begin{equation}
4\pi \al \la^2 - \al
\langle 0 |E\mA^{-1}\Ea|0\rangle = 8\pi\al [\la -\ln(1+\la)]
\end{equation}
we infer
\begin{equation*}
(\Psi, \bh \Psi) /(\Psi,\Psi) = -e_0 + 8\pi\al [\la -\ln(1+\la)] -4 \al
(\Da p_x \phi, A_V^{-1}\Da p_x \phi)+ \Ow(\al^2),
\end{equation*}
which proves the first statement of the Theorem with
\begin{equation}\label{rc}
\Em(V,\la) = 4 \al
(\Da p_x \phi, A_V^{-1}\Da p_x \phi).
\end{equation}

The second statement follows by the observation that  $(\Da p_x \phi, A_V^{-1}\Da p_x
\phi)$ is strictly larger than $0$, which is a consequence of the fact that 
$\Da p_x \phi $ is a not identically vanishing function $\in \Ll^2(\R^3 \otimes (\R^3,\C^2))$ and $A_V$ an invertible operator.

The statement for $\la >\bar \la$ follows from \cite[Theorem 2.1]{H1} and above estimates.

\section{Computation of concrete numbers}\label{shift}

\subsection{Error for the self-energy}

We are going to calculate the error $\err$ of the self-energy,
\begin{equation}
|\so - 8\pi \al[\la - \ln(1+\la)]| \leq \err(\al^2).
\end{equation}
For sake of convenience, we treat only the case $\la \leq \bar \la$.
The most important part is to estimate the kinetic energy term $\pa p\Psi\pa^2$ 
for $\Psi$ close to the ground state $\so$.

By means of (\ref{39}), (\ref{310}) and then applying (\ref{energy}) we obtain
\begin{eqnarray*}
8\pi \al \la &\geq& (\Psi,T\Psi) \geq (1-a) \pa p\Psi\pa^2 + (\Psi,H_f\Psi) -\\ &&- 4a^{-1}\al\pa D\Psi\pa^2 -c\al \la^3 - 1/(c\la^3) \pa E\Psi\pa^2
\\ &\geq& (1-a)\pa p\Psi\pa^2 - c \al\la^3 +\left[ 1 - a^{-1} 32 \pi \al \la - c^{-1} \frac{8\pi}3\right](\Psi,H_f\Psi),
\end{eqnarray*}
where $ 1 > a > 0$.
We require the last term $[..]$ to be $\geq 0$. For simplicity, we choose $c =8\pi$, then our first condition on $\al$ reads
\begin{equation}\label{req}
\al \leq \frac a{16\pi\la}
\end{equation}
and additionally
\begin{equation}\label{ppsi}
\pa p\Psi\pa^2 \leq \frac{8\pi\al\la(1+\la^2)}{1-a}.
\end{equation}

It is not difficult to see that the main contribution to $\err$ stems from (\ref{smc}) and (\ref{mc}). 
When we take that  value twice we obtain a rather accurate upper bound for $\err $.
From the Section 3.2 we know that (\ref{mc}) is bounded by
\begin{equation*}
8\pi \al \la \pa p\Psi\pa^2 + 8\pi \al \la^2 (\Psi, H_f\Psi).
\end{equation*}
From the proof of Theorem 1 in \cite{H1} we infer 
\begin{equation*}
(\ref{smc}) \leq 6\pi \al \la^2  (\Psi, H_f\Psi),
\end{equation*}
which implies, after doubling the value,
\begin{eqnarray*}
\err &\leq& 16\pi\al  \la \pa p\Psi\pa^2 + 28\pi \al \la^2 (\Psi, H_f\Psi)\\
&\leq& \al^2 \left[ \frac{(16\pi)^2\la^2(1+\la^2)}{1-a} + 14\cdot 16\pi^2 \la^2\right]
\end{eqnarray*}
where we used (\ref{ppsi}) and (\ref{aprih}).

\subsection{Electromagnetic Energy-shift}

We consider an electron in the field of a nucleus with charge $Z$, i.e.
\begin{equation*}
V = - \frac {Z\beta}{|x|},
\end{equation*}
where $\be$ is the \lq\lq real\rq\rq \ fine structure constant $\be = 1/137$.
The ground state energy of the corresponding Schr\"odinger operator $p^2+ V$ is known to be
\begin{equation*}
\infspec [p^2+ V] = -e_0 = - \frac 14 (\be Z)^2 
\end{equation*}
measured in units $[\frac {m_ec^2}{2}]$ with $m_e$ the electron mass. The corresponding radiative correction,  obtained in (\ref{rc}), is given by 
\begin{equation}\label{radcorr}
\Em(V,\la) = - \al 4 (\phi,pD A_V^{-1} p\Da \phi),
\end{equation}
where $\phi=\phi(|x|)$ denotes the ground state of $p^2 + V$.
We know
\begin{equation}\label{phiz}
\nabla \phi(|x|) = \partial_r \phi(r) \vec e_r(\theta, \varphi)= e_0^{1/2} \phi(r)  \vec e_r(\theta, \varphi)
\end{equation}
when using polar coordinates.
Denote with $\phi_i$ the eigenstate of $p^2+ V$ with corresponding  eigenvalue $-e_i$.
Then, by means of (\ref{phiz}), we obtain, by straightforward computations,
\begin{equation}
(\phi,pD A_V^{-1} p\Da \phi) = 4\pi e_0\sum_{i\geq 1} |c_i|^2 \int_0^\la\frac{p \Theta(\la - p)}{e_0 - e_i + p^2 + p} dp \equiv e_0 F(\la),
\end{equation}
where
\begin{equation*}
c_i = \int \overline{\phi_i(r,\theta,\varphi)} \phi(r) \cos(\theta) r^2 dr d \Omega,
\end{equation*}
with $d \Omega = sin \theta d\theta d\varphi$.
Notice, $\sum |c_i|^2 = 2/3$ and 
\begin{equation*}
\frac{8\pi}3\ln[1+\la] /F(\la) \to 1
\end{equation*}
as $\la \to \infty$. That is why, for simplicity, we take $\frac{8\pi}3\ln[1+\la]$ instead of $F(\la)$
and obtain an approximative {\it radiative correction} 
\begin{equation}\label{hain}
R_C = \al e_0\frac{32\pi}3  \ln[1+\la] 
\end{equation}
for the binding energy. (Indeed, for $\la \sim 1$ these two functions perfectly coincide, but still for $\la=e_0$,
$F(\la)/\frac{8\pi}3\ln[1+\la] \sim 0.3$.) 

Notice that in contrary to Bethe's formula in \cite{Be}, 
where the energy shift grows with $Z^4$, our correction term 
only increases as $Z^2$.

Consider a hydrogen or helium atom and  take $\la =2$ then (\ref{hain}) turns out to be {\it much} too large compared with
experiment, whereas setting $\la \sim e_0$ yields  a correction
$R_C \sim \be (\be Z)^4 \sim \be^3  Z^4$Ry, which agrees with Bethe's prediction (if $\be Z \ll 1$) and ranges within experimental data.

\subsection{Calculating concrete values of $\al$ and $\la$}

We search for values of $\al$ and $\la$, that guarantee  the error of the self energy $\err$ being smaller than the 
radiative correction $R_C$. This leads to the condition:
\begin{equation*}
\al \leq {\rm Min} \left\{ \frac{\frac 23 e_0 \ln[1+\la]}{\frac{16\pi \la^2[1 +\la^2]}{1-a} + 14 \pi \la^2}, \frac a{16\pi\la}\right\}. 
\end{equation*}
Set $\la = e_0 \ll 1$ $(Z \sim 1)$ and take into account $\ln(1 + e_0) \sim e_0$
then for
\begin{equation*}
\al \leq 1/45\pi
\end{equation*}
enhanced binding is guaranteed.
(Recall, this bound is by far not optimal, since for small $\la$ we are able to work
with much better a priori estimates, $(\Psi,T\Psi) \lesssim 8\pi \al \la^2$, from the beginning, which would
imply an $\al \lesssim  1/e_0 45\pi\sim 100$.)

Now we take $\la =1$ and $a=1/4$. Then
\begin{equation*}
\al \leq {\rm Min} \left\{e_0/21,1/200\right\}.
\end{equation*}
For $Z\be=1$ the $\al$ required  is $1/200$,
which still  coincides quite well with the physical value,
whereas for $Z=2$ we need $\al \leq 1/(4\cdot 10^5)$ to dominate $\err$ by $R_C$.

We conclude that in the case of $\la \sim e_0$, which seems to be reasonable,
we obtain \\
(1) an energy shift comparable to experiment,\\
and\\
(2) the error of the self-energy, in the case of $\al = 1/137$, is (much) less than
$R_C$.

\subsection{A comment on mass renormalization}

Lets re-introduce for the moment the electron mass in the definition of $T$ and $p^2$.
One way of defining the renormalized mass (cf. \cite{LL2}) is to set the binding energy
equal to the physical binding energy, i.e.
\begin{equation*}
\half m_{\rm phys} (Z\be)^2 \equiv \so - E_\al, 
\end{equation*}
which, in the case of small $\al$ and assuming that (\ref{es}) holds with equality,
implies
\begin{equation*}
 m_{\rm phys} =m\left[1 + \al 4 F(\la/m)\right]. 
\end{equation*}

For $\la/m \sim 1$ this perfectly agrees
with
\begin{equation*}
 m_{\rm phys} =m\left[1 + \al \frac{32\pi}3 \ln(1 +\la/m)\right], 
\end{equation*}
which is exactly what we obtain (for $\al \to 0$), when defining the renormalized mass
by means of the ground state energy of a free electron with fixed total (electron + photon)
momentum $P$.

\end{document}